\documentclass[aps,preprintnumbers,amsmath,amssymb,showpacs,nofootinbib,onecolumn,superscriptaddress]{revtex4}
\usepackage{amsfonts}
\usepackage{multirow}
\usepackage{makecell}
\usepackage{mathrsfs}
\usepackage{graphicx}
\usepackage{amsmath}
\usepackage{amssymb}
\usepackage{bm}
\usepackage{bbm}
\usepackage{color}
\usepackage{ulem}
\usepackage{array}
\usepackage{diagbox}
\usepackage{float}
\allowdisplaybreaks[4] %%%Allow long expressions displayed with break across the pages
\def\OMIT#1{}

\def\hlinew#1{%
  \noalign{\ifnum0=`}\fi\hrule \@height #1 \futurelet
   \reserved@a\@xhline}
\usepackage{array}
\newcommand{\PreserveBackslash}[1]{\let\temp=\\#1\let\\=\temp}
\newcolumntype{C}[1]{>{\PreserveBackslash\centering}p{#1}}
\newcolumntype{R}[1]{>{\PreserveBackslash\raggedleft}p{#1}}
\newcolumntype{L}[1]{>{\PreserveBackslash\raggedright}p{#1}}

\newcommand{\nn}{\nonumber}

\newcommand{\beq}{\begin{equation}}
\newcommand{\eeq}{\end{equation}}
\newcommand{\bqa}{\begin{eqnarray}}
\newcommand{\eqa}{\end{eqnarray}}

\newcommand\fverb{\setbox\fverbbox=\hbox\bgroup\verb}
\newcommand\fverbdo{\egroup\medskip\noindent%
            \fbox{\unhbox\fverbbox}\ }
\newcommand\fverbit{\egroup\item[\fbox{\unhbox\fverbbox}]}
\newbox\fverbbox

\makeatletter

\newcommand{\Rmnum}[1]{\expandafter\@slowromancap\romannumeral #1@}
\allowdisplaybreaks[2]

\makeatother

\begin{document}

%\preprint{APS/123-QED}

\title{Two loop QCD corrections to $e^+ e^- \to J/\psi + \eta_c$ in asymptotic expansion}% Force line breaks with \\

%\author{}%
%\affiliation{
%}%

\author{Cong Li}%~\footnote{lc312321@163.com}
\affiliation{School of Physical Science and Technology, Southwest University, Chongqing 400700, China\vspace{0.2cm}}

\author{Xu-Dong Huang}%~\footnote{huangxd@cqnu.edu.cn}
\affiliation{College of Physics and Electronic Engineering, Chongqing Normal University, Chongqing 401331, China\vspace{0.2cm}}

\author{Wen-Long Sang~\footnote{wlsang@swu.edu.cn}}
\affiliation{School of Physical Science and Technology, Southwest University, Chongqing 400700, China\vspace{0.2cm}}

\date{\today}

%-------------------------------------------------
\begin{abstract}
Within the framework of NRQCD, the two-loop QCD corrections to the short-distance coefficient for the process $e^+e^-\to J/\psi+\eta_c$ have been obtained in asymptotic expansion over $r={16m_c^2}/{s}$ up to $r^{15}$, sufficient for phenomenological studies. The obtained leading logarithmic terms can facilitate the resummation of large logarithms.  We provide cross section predictions in both on-shell and $\overline{\rm MS}$ mass schemes, including $\mu_R$ uncertainties. Both mass schemes give similar results, consistent with experimental data.
\end{abstract}
%--------------------------------------------------

%\pacs{12.38.Bx, 12.39.St, 13.85.Ni, 14.40.Pq}% PACS, the Physics and Astronomy
%\keywords{Suggested keywords}%Use showqkeys class option if keyword
\maketitle
%------------------------------------------------
%{\it Introduction}
%-----------------------------------------------
 \section{Introduction} 
The production of double charmonium states in electron-positron annihilation, exemplified by the process $e^+e^- \to J/\psi+\eta_c$, serves as an important platform to explore the perturbative and non-perturbative nature of QCD. The process $e^+e^- \to J/\psi + \eta_c$ was first measured by the {\tt BELLE} collaboration, yielding $\sigma_{e^+e^- \to J/\psi+\eta_c} \times B_{\geq4} = 33^{+7}_{-6} \pm 9$ fb~\cite{Abe:2002rb}, which was later revised to $ \sigma_{e^+e^- \to J/\psi+\eta_c} \times B_{>2} = 25.6 \pm 2.8 \pm 3.4 $ fb ~\cite{Abe:2004ww}. Here $ B_{>n} $ denotes the branching fraction of $ \eta_c $ decaying into $n$ charged tracks. In 2005, an independent measurement by the {\tt BABAR} collaboration reported $\sigma_{e^+e^- \to J/\psi+\eta_c} \times B_{>2} = 17.6 \pm 2.8^{+1.5}_{-2.1}$ fb ~\cite{Aubert:2005tj}.
Recent searches by {\tt BELLE} at $\Upsilon(nS)$ resonances $(n = 1–5)$ and $10.52$ GeV off-resonance energy found no significant signal~\cite{Belle:2023gln}.

These experimental measurements have sparked a flurry of theoretical investigations in the following years~\cite{Braaten:2002fi,Liu:2002wq,Brodsky:2003hv,Hagiwara:2003cw,Cheung:2003xw,Ma:2004qf,Bondar:2004sv,Liu:2004ga,Zhang:2005cha,Braguta:2005kr,Bodwin:2006dm,Ebert:2006xq,He:2007te,Choi:2007ze,Bodwin:2007ga,Gong:2007db,Guo:2008cf,Mengesha:2011pu,Dong:2012xx,Li:2013otv,Bodwin:2014dqa,Sun:2018rgx,Feng:2019zmt,Zeng:2021hwt,Sun:2021tma,Huang:2022dfw}.
The NRQCD factorization~\cite{Bodwin:1994jh} has emerged as a cornerstone for describing the process $e^+e^- \to J/\psi + \eta_c$. Leading-order (LO) NRQCD predictions were initially carried out in Refs.~\cite{Braaten:2002fi,Hagiwara:2003cw,Liu:2002wq,Liu:2004ga}, but these were found to be an order of magnitude smaller than the experimental measurements. An encouraging improvement came with the calculation of the next-to-leading order (NLO) radiative corrections, which were substantial and positive, significantly reducing the discrepancy between theory and experiment~\cite{Zhang:2005cha,Gong:2007db}. Meanwhile, both $\mathcal{O}(v^2)$ and $\mathcal{O}(\alpha_sv^2)$ corrections were also explored~\cite{Braaten:2002fi,He:2007te,Dong:2012xx,Li:2013otv}.

Recently, the next-to-next-to-leading-order (NNLO) perturbative corrections were evaluated in Refs.~\cite{Huang:2022dfw, Feng:2019zmt}. The analysis of Ref.~\cite{Feng:2019zmt} reveals that although the $\mathcal{O}(\alpha_s^2)$ correction is smaller than the $\mathcal{O}(\alpha_s)$ term, both are comparable in size to the LO contribution—a pattern that signals a slow perturbative convergence.  While the inclusion of all radiative and relativistic corrections brings the theoretical prediction into consistency with the experimental measurement~\cite{Huang:2022dfw}, significant uncertainties remain from the renormalization scale and the charm quark mass, which undermine the clarity of this agreement. 

More seriously, it has been noted that the renormalized pole mass is subject to a non-perturbative renormalon ambiguity of $\mathcal{O}(\Lambda_{\rm QCD})$~\cite{Beneke:1994sw,Bigi:1994em,Beneke:1994rs}, which not only affects the definition of the pole mass but also potentially impacts the perturbative expansion of the cross section. A partial resolution of this issue can be achieved by converting the cross section from the on‑shell (OS) mass scheme to the $\overline{\rm MS}$ mass scheme, where the latter is expected to be free from such ambiguities. Additionally, it has been found that the leading logarithmic term $\ln^2 \frac{s}{m^2}$ dominates the NLO radiative corrections~\cite{Jia:2010fw,Bodwin:2014dqa}. Whether large logarithmic terms will also dominate the NNLO corrections remains unknown. If an analytic expression can be derived, it will not only address this question but also facilitate the resummation of large logarithms. Moreover, the analytic expression will greatly facilitate phenomenological applications, particularly when different center-of-mass (CM) energies $\sqrt{s}$ or charm quark masses are considered.

Unfortunately, it seems insurmountable to derive the analytic expression for the cross section of $e^+e^- \to J/\psi+\eta_c$ due to the extremely complicated topology in the Feynman diagrams. Thus, our task in this work is to obtain the asymptotic expansion over $r=16m_c^2/s$. This asymptotic expression includes various logarithmic terms and provides a good approximation of the exact results over a wide range of $r$. Therefore, it allows us to easily convert the theoretical prediction from the OS mass scheme to the $\overline{\rm MS}$ mass scheme and make various phenomenological applications.

%NRQCD factorization for the FF and cross section
\label{sec-nrqcd}
\section{NRQCD factorization}
The cross section for the process $e^+e^- \to J/\psi + \eta_c$ can be expressed in terms of the time-like electromagnetic (EM) form factor $F(s)$, which is defined via 
%------------------------------------
\beq
%------------------------------------
\langle J/\psi(P_1,\lambda)+\eta_c(P_2)\vert J_{\rm EM}^\mu \vert 0
\rangle = i\,F (s)\,\epsilon^{\mu\nu\rho\sigma} P_{1\nu} P_{2 \rho}
\varepsilon^*_\sigma (\lambda),
%------------------------------------
\label{EM:EM}
%------------------------------------
\eeq
%------------------------------------
where $J_{\rm EM}^\mu$ denotes the EM current, $P_1$ and $P_2$ are the momenta of the $J/\psi$ and $\eta_c$ respectively, $s=(P_1+P_2)^2$, and $\varepsilon$ represents the polarization vector of $J/\psi$ with $\lambda=\pm 1 ,0$ indicating its helicties.

By employing NRQCD factorization~\cite{Bodwin:1994jh}, $F(s)$ can be expressed as:
%------------------------------------
\beq
%------------------------------------
F(s)=\sqrt{2 M_{J/\psi}} \sqrt{2 M_{\eta_c}} \frac{\langle J/\psi| \psi^\dagger
\bm{\sigma} \cdot \bm{\epsilon} \chi |0\rangle}{m_c} \frac{\langle \eta_c|
\psi^\dagger \chi | 0 \rangle}{m_c} \times f,
%------------------------------------
\label{EM:NRQCD}
%------------------------------------
\eeq
%------------------------------------
where the short-distance coefficient (SDC) $f$ encodes the perturbatively calculable physics, while non-perturbative long-distance 
effects are described by long-distance matrix elements (LDMEs).
The prefactor $\sqrt{2 M_{J/\psi}} \sqrt{2 M_{\eta_c}}$ arises from the relativistic normalization of quarkonium states in the helicity amplitude and their non-relativistic normalization in LDMEs. Note that the charm quark mass $m_c$, which originates from the perturbative matching, has been explicitly factored out in \eqref{EM:NRQCD}.

The cross section is then readily obtained:
%-----------------------------
\bqa
%-----------------------------
\sigma[e^+e^-\to J/\psi + \eta_c] = \dfrac{4\pi \alpha^2}{3}
\left(\dfrac{|{\bm P_{1}}|}{\sqrt{s}}\right)^3  \left|F(s)\right|^2\nn\\
=\dfrac{16\pi\alpha^2 M_{J/\psi} M_{\eta_c}|{\bm P_{1}|}^3 }{3s^\frac{3}{2}}\frac{\langle
{\mathcal O}\rangle_{J/\psi}}{m_c^2} \frac{ \langle{\mathcal
O}\rangle_{\eta_c}}{m_c^2} \,|f|^2,
%---------------------------------
\label{cross:section}
%---------------------------------
\eqa
%---------------------------------
where $|{\bm P_{1}}|$ is the magnitude of the spacelike momentum carried by $P_1$, and we have used the shorthand notations $\langle \mathcal{O} \rangle_{J/\psi} = \big| \langle J/\psi \vert \psi^\dagger \bm{\sigma} \cdot \bm{\epsilon}_{J/\psi} \chi \vert 0 \rangle \big|^2$ and $\langle \mathcal{O} \rangle_{\eta_c} = \big| \langle \eta_c \vert \psi^\dagger \chi \vert 0 \rangle \big|^2$.

Note that we retain the physical masses of the  $J/\psi$ and $\eta_c$ in the cross section \eqref{cross:section}, which originate from the normalization of the states, phase space factors, or the Lorentz contraction from~\eqref{EM:EM}. In contrast, in the computation of $f$, we take the masses of the $J/\psi$ and $\eta_c$ to be $2m_c$.

We expand the SDC $f$ in powers of $\alpha_s$:
%-----------------------------
\bqa
%-----------------------------
f(r)&=&f^{(0)}(r)
\Big[1+\frac{\alpha_s}{\pi} \left(\frac{\beta_0}{4} \ln \frac{4\mu_R^2}{s} +f^{(1)}(r)\right)\nn\\
&+&\frac{\alpha_s^2}{\pi^2} \Big(\frac{\beta_0^2}{16} \ln^2 \frac{4\mu_R^2}{s}+\big(\frac{\beta_1}{16}+\frac{\beta_0}{2}f^{(1)}(r)\big) \ln \frac{4\mu_R^2}{s}\nn\\
&+&\big(\gamma_{J/\psi}+\gamma_{\eta_c}\big) \ln \frac{\mu_\Lambda^2}{m_{c}^{2}}+f^{(2)}(r)\Big)+\mathcal{O}\left(\alpha_s^3\right)\Big],
%-----------------------------
\label{eq-sdcs}
%-----------------------------
\eqa
%-----------------------------
where $f^{(0)}$ denotes the LO SDC, $\beta_0 = \frac{11}{3} C_A - \frac{4}{3} T_f n_f$ and $\beta_1=\frac{34}{3}C_{A}^{2}-\frac{20}{3}C_A T_f n_f -4C_F T_f n_f$ are 
the one-loop  and two-loop coefficients of the QCD $\beta$ function respectively. The number of active flavors is taken as $n_f=n_l+n_h$, with $n_l$=3 being the number of light quarks, and $n_h=1$ the number of heavy quark. $\mu_R$ and $\mu_\Lambda$ refer to the renormalization scale and  the NRQCD factorization scale, respectively. The anomalous dimensions $\gamma_{J/\psi}$ and $\gamma_{\eta_c}$ associated with the NRQCD bilinear currents carrying the quantum numbers $^3S_1$ and $^1S_0$ are given by Refs.~\cite{Czarnecki:1997vz,Beneke:1997jm,Czarnecki:2001zc}.

%{\it theoretical framework and computational methods}
\label{sec-tech}
\section{Computational methods}
\begin{figure}[t]
\begin{center}
\includegraphics[width=0.8\textwidth]{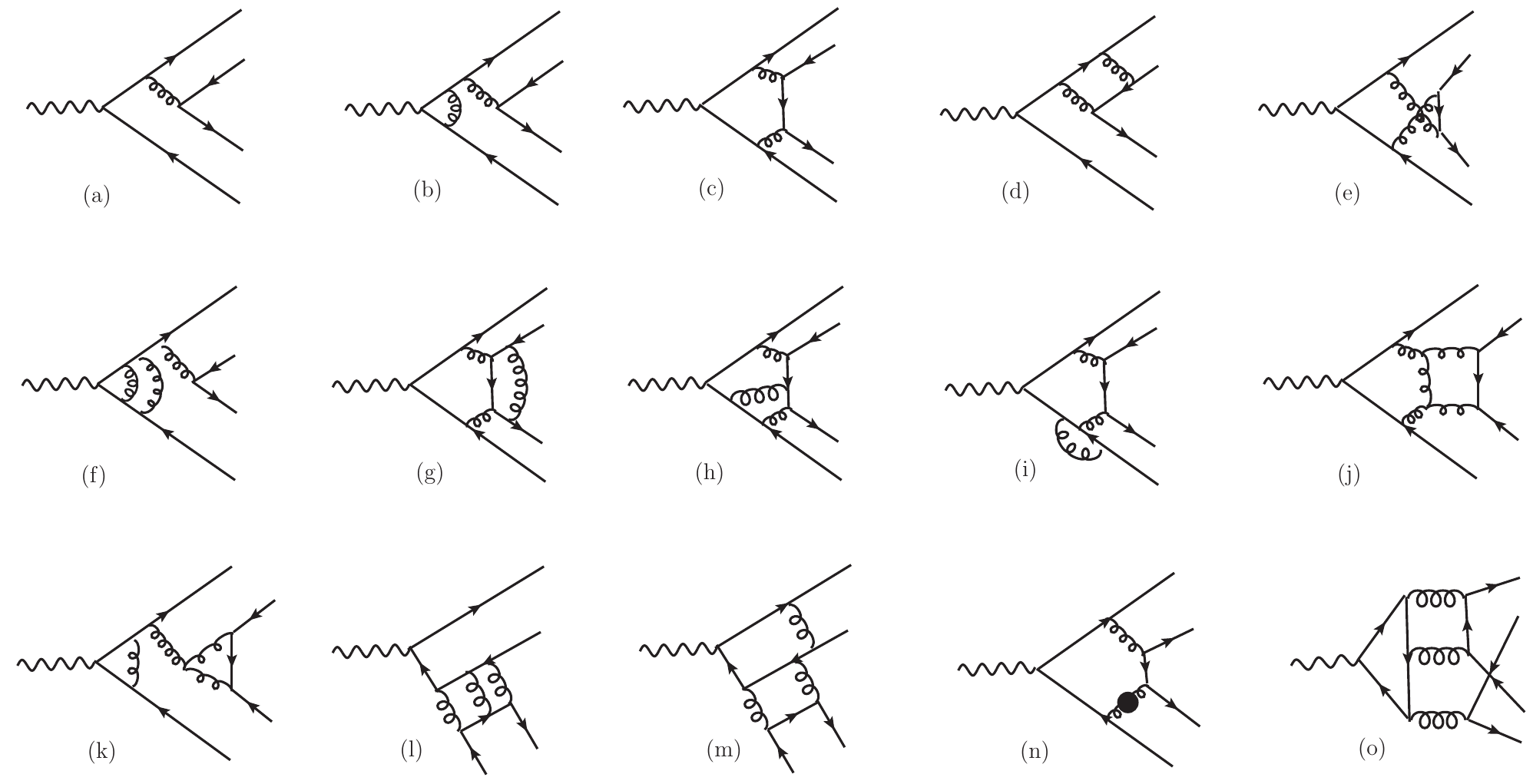}
\end{center}
\caption{Some typical Feynman diagrams. (a) is the LO amplitude, (b)-(e) illustrate the NLO corrections, and (f)-(o) depict the NNLO contributions.  The final diagram (o) illustrates the “light-by-light” part.}
\label{feynman}
\end{figure}

We utilize {\tt FeynArts}~\cite{Hahn:2000kx} to generate the Feynman diagrams and amplitudes for the process $\gamma^{*} \to c \bar{c}(^{3} S_{1}) + c \bar{c}(^{1} S_{0})$, with representative diagrams depicted in Fig.~\ref{feynman}. The packages {\tt FeynCalc} and {\tt FormLink}~\cite{Mertig:1990an,Feng:2012tk} are employed to handle Lorentz contractions and Dirac algebra. To LO in $v$, we neglect the relative momentum within each $c\bar{c}$ pair before performing the loop integration, allowing us to directly extract the NRQCD SDCs from the hard region~\cite{Beneke:1997zp}. The Feynman integrals are classified using {\tt CalcLoop}~\cite{CalcLoop}.
We use {\tt Kira}~\cite{Klappert:2020nbg} and  {\tt Blade}~\cite{Guan:2024byi} to reduce the scalar integrals to the master integrals (MIs), and then we compare the results to cross-check the reductions. We further used {\tt FIRE}~\cite{Smirnov:2014hma} to look for the symmetry between different families and MIs. We work in $d=4-2\epsilon$ spacetime dimensions to regularize both UV and IR divergences.

%-------------------------------
%-------------------------------

The asymptotic expansions of the MIs as $r\to 0$ are derived via differential equations (DEs) with respect to the kinematic variable $r$. For a complete set of MIs $\vec{J}$, the DEs take the form:
%-----------------------------
\bqa
%-----------------------------
\frac{\partial }{\partial r}\vec{J}=M(r,\epsilon) \vec{J},
%-----------------------------
\label{eq-de}
%-----------------------------
\eqa
%-----------------------------
where the matrix $M$ is a local Fuchsian form at $r=0$. The solution for $\vec{J}$ can be expressed as a series expansion in $r$:
%-----------------------------
\bqa
%-----------------------------
\vec{J}=\sum_{\rho \in S} r^\rho \sum_{i=0}^{\infty}\vec{C}_{\rho,i}(\epsilon) r^i,
%-----------------------------
\label{eq-asy}
%-----------------------------
\eqa
%-----------------------------
with $S$ being a finite set of $\epsilon$-dependent exponents of the form $-j+k\,\epsilon$, where $j$ and $k$ are rational numbers. Substituting \eqref{eq-asy} into \eqref{eq-de}, we are able to determine the exponents $\rho$ by matching the lowest-order terms and compute the coefficients $\vec{C}$ from recurrence relations and initial conditions.

In our practical computations, we employ a numerical fitting strategy. 
We first solve the DEs for a set of $\epsilon$ values ($\epsilon_0, ..., \epsilon_n$), then determine the exponents $\rho$ and the coefficients $\vec{C}$ as expansions of $\epsilon$ 
%-----------------------------
\bqa
%-----------------------------
\vec{a}_0 +\vec{a}_1 \epsilon+...+\vec{a}_n \epsilon^n
%-----------------------------
%-----------------------------
\eqa
%-----------------------------
by solving systems of linear equations. This approach is found to be very efficient and is implemented with the auxiliary mass flow package~\cite{Liu:2017jxz,Liu:2022mfb,Liu:2022chg}.

Once the MIs are known, we further expand both $\rho$ and $\vec{C}$ in $\epsilon$. Each integral is thus expressed as a series in $\epsilon$ and $r$, and the final result takes the form $\sum_{i,j} f_{i,j} r^i\ln^j r$ for each $\epsilon$ order. 
In this work, we truncate the expansions in $r$ up to $r^{15}$,  which is sufficient to accurately reproduce the exact results for a wide range of $r$. Given the high precision of the coefficients $f_{i,j}$,  we use the {\tt PSLQ} algorithm to successfully reconstruct the analytic expressions of  $f_{i,j}$ up to $\ln^2 r$  for some lower powers of $r$. 

In this study, the OS scheme is employed to renormalize the charm quark mass and field strength, while the $ \overline{\text{MS}} $ scheme is utilized for the QCD coupling constant. The renormalized NNLO amplitude retains a single IR pole. The coefficient of this pole exactly matches half the sum of the anomalous dimensions for the NRQCD bilinear operators that carry the quantum numbers of $ J/\psi $ ~\cite{Czarnecki:1997vz,Beneke:1997jm} and $ \eta_c $ ~\cite{Czarnecki:2001zc}. Under the $ \overline{\rm{MS}} $ scheme, this IR pole factorizes into the corresponding NRQCD matrix elements.

After performing a series of intricate calculations, we have successfully derived the expressions for $f^{(n)}$ with $n=0,1,2$.

The LO SDC has long been established in the literature~\cite{Braaten:2002fi}:
%---------------------
\begin{eqnarray}\label{eq:LO}
%---------------------
f^{(0)}=\frac{64C_F\pi\alpha_s m_c}{9 s^2},
%---------------------
\end{eqnarray}
%---------------------
where $m_c$ arises due to the helicity flip in the process $e^+e^-\to J/\psi+\eta_c$.

The $f^{(1)}(r)$ and $f^{(2)}(r)$ are expressed in an expansion over $r$:
%---------------------
\bqa
%---------------------
f^{(n)}(r) = \sum_{i,j\ge 0} f^{(n)}_{i,j} r^i \ln^j r,
%---------------------
\eqa
%---------------------
where $j$ ranges from $0$ to $2n$, and $i$ is truncated up to $15$. 
Specifically, the leading logarithmic coefficients are:
%---------------------------------
\bqa\label{f02n}
%---------------------------------
f^{(1)}_{0,2}&=&\frac{11C_F}{16}-\frac{C_A}{8},\nn \\
f^{(2)}_{0,4}&=&\frac{19C_{F}^{2}}{384}-\frac{C_A C_F}{64}.
\eqa
%---------------------------------
The remaining $\mathcal{O}(r^0)$ terms at NLO read
%-----------------------------------------------
\begin{eqnarray}\label{one-ex1}
%---------------------------------
f^{(1)}_{0,1}&=&-\frac{41}{24}-\frac{11 \ln2}{12}+\frac{13 i \pi}{12},\nn \\
%---------------------------------
f^{(1)}_{0,0}&=&-\frac {1} {6} + \frac {277\ln 2} {24} - \frac {61 \ln^2 2} {24} - \frac {\pi^2} {36} \nn\\
&+& i\left(\frac {25\pi } {24} - \frac {11\pi \ln 2} {12}\right)-\left(\frac {5 } {18} + \frac {i\pi} {6} \right)n_h\nn ,
%---------------------------------
\end{eqnarray}
%---------------------------------
while those at NNLO are
%------------------------------------
\begin{eqnarray}\label{two-ex1}
%---------------------------------
f^{(2)}_{0,3}&=&-\frac{275}{144}+\frac{419 \ln2}{864}+\frac{11 i\pi}{108}+\frac{13n_h}{216}+\frac{13n_l}{216},\nn \\
%---------------------------------
f^{(2)}_{0,2}&=&\frac {3943} {576} + \frac {2695\ln 2} {288} - \frac
{1039\ln^2 2} {288} - \frac {173\pi^2} {864}\nn\\
&-& i\left(\frac {11 \pi} {4} - \frac {419\pi\ln 2} {288}\right)-\left(\frac {787} {1728}+\frac {37 \ln 2} {144}\right) n_h\nn \\
&-&\left(\frac {787} {1728} + \frac {37\ln 2} {144}\right) n_l-\left(\frac {5} {192} - \frac {5\ln^2 2} {96} \right)_{lbl},\nn \\
%---------------------------------
f^{(2)}_{0,1}&=&-(60.3367 -26.2776 i)+(4.8134 - 1.5263 i) n_h \nn\\&+& (4.8134 -1.5263 i) n_l -\left(0.0148 + 0.0064 i\right)_{lbl},\nn\\
%---------------------------------
f^{(2)}_{0,0}&=&-(7.7964 - 31.0330 i) - (0.8070 + 6.2668 i)n_h\nn\\
&-& (7.0174 + 6.2668 i)n_l 
-(0.3940 - 0.5818 i) n_h n_l \nn\\
&-& (0.1970 - 0.2909 i) n_h^2 - (0.1970 - 0.2909 i)n_l^2 \nn\\
&-& (0.0732 + 0.0466 i)_{lbl},
%---------------------------------
\end{eqnarray}
%---------------------------------
 where the subscript {\it lbl} indicates the contribution from the ``light-by-light" Feynman diagrams.

The asymptotic expressions for $f^{(1)}(r)$ and $f^{(2)}(r)$ up to $r^{15}$, 
along with the leading logarithmic coefficients for each Feynman diagram, are included in the attached file. For convenience, the analytic expression for $f^{(1)}(r)$ and the numerical results for $f^{(2)}(r)$ at $619$ different values of $r$ ranging from $0$ to $1$  are also provided in the attached file. The latter is calculated using a strategy analogous to that described in Refs.~\cite{Feng:2019zmt,Sang:2022kub,Sang:2023liy}. 

%{\it $\overline{\text{MS}}$ mass scheme}
\section{$\overline{\text{MS}}$ mass scheme\label{sec-mass-scheme}}
As previously discussed, the $\overline{\text{MS}}$ mass is expected to be free from non-perturbative ambiguities. In fact, converting $f(r)$ from the OS mass scheme to the $\overline{\text{MS}}$ mass scheme is straightforward. This can be achieved by making the replacement 
%------------------------------------
\begin{equation}\label{eq:ms:os:rel}
%------------------------------------
m_c = \overline{m}_c(\mu_m)\left(1 + \sum_{n=1}^\infty \left(\frac{\alpha_s(\mu_m)}{\pi}\right)^n d^{(n)}(\mu_m)\right),
%------------------------------------
\end{equation}
%------------------------------------
and truncating the expansion at $\mathcal{O}(\alpha_s^2)$. Here $\overline{m}_c(\mu_m)$ is the $\overline{\text{MS}}$ mass at $\mu_m$ and the coefficients $d^{(n)}(\mu_m)$ are known through four loops~\cite{Gray:1990yh,Fleischer:1998dw,Chetyrkin:1999qi,Melnikov:2000qh,Marquard:2016dcn}. It is worth noting that, while $\mu_m$  is formally arbitrary in principle, it is convenient to assign the same central values to both $\mu_m$ and $\mu_R$, and then estimate the uncertainties by considering variations around these central values.

It is also important to note that we need to convert the $m_c$ that explicitly appears in Eq.~\eqref{cross:section} and stems from the perturbative matching in the rest frame of the charmonium states. Clearly, it is not reasonable to choose $\mu_m$ to be related to the typical collision energy $\sqrt{s}$. Instead, it is natural to choose $\mu_m$ at $\mathcal{O}(m_c)$. In our phenomenological analysis, we fix $\mu_m = \overline{m}_c(\overline{m}_c)$ when converting this specific $m_c$ to the $\overline{\mathrm{MS}}$ mass $\overline{m}_c(\mu_m)$ via Eq.~\eqref{eq:ms:os:rel}, while elsewhere we simply take $\mu_m = \mu_R$.

\label{sec-convergence}
\section{Convergent behaviour of the asymptotic expansion}
To facilitate this analysis, we first introduce a shorthand notation:
%-----------------------------
\bqa
%-----------------------------
g^{(2)}_{a,b} = \sum_{i=0}^{a}\sum_{j=b}^{4}f^{(2)}_{i,j}r^i\ln^{j}r.
%-----------------------------
\eqa
%-----------------------------
Additionally, we use $g^{(2)}_{\rm{tot}}$ to represent the exact value of $f^{(2)}$.

\begin{figure}[!h]
\hspace{0cm}\includegraphics[width=0.45\textwidth]{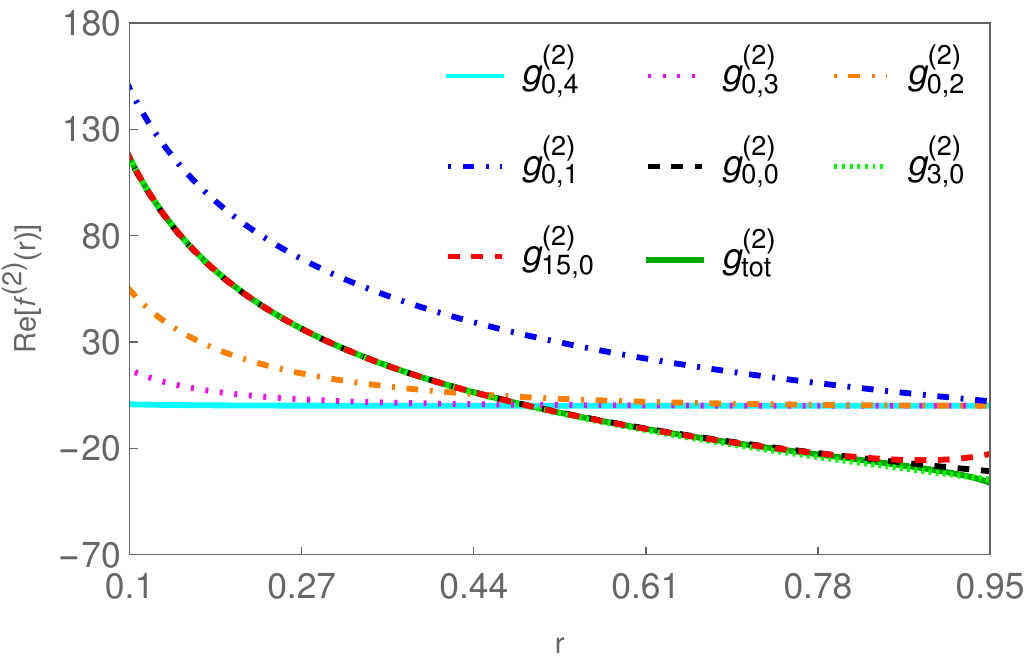}
\hspace{0cm}\includegraphics[width=0.45\textwidth]{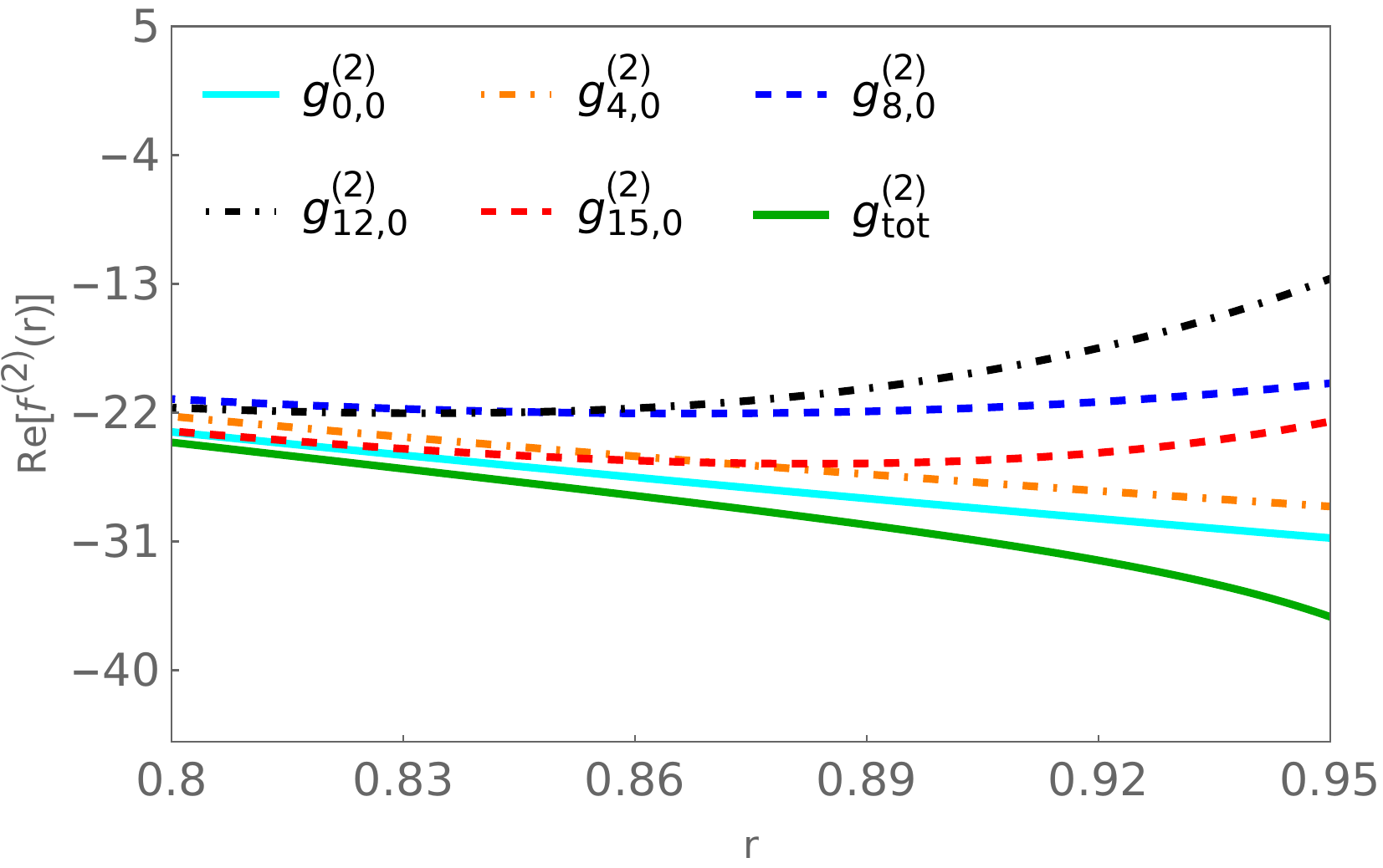}
\caption{\label{fig:loop2:re}
Comparison of the asymptotic expressions with the exact results for real part of $f^{(2)}(r)$.}
\end{figure}

Figure~\ref{fig:loop2:re} compares the real part of $f^{(2)}$  between the asymptotic expressions and the exact results. Although the contribution from the leading logarithmic term $r^0\ln^4r$ is significantly smaller than the exact result, the asymptotic expression up to $r^0$ already provides a reasonable approximation to the exact result for $r<0.8$. 
However, the convergence deteriorates rapidly for $r>0.8$. While higher-order terms in the asymptotic expansion improve the convergence for smaller $r$, they do not enhance it for larger $r$.~\footnote{ It is worth noting that including the asymptotic expansion up to $r^{30}$ only slightly reduces the deviation from the exact result for $r<0.8$, yet enlarges it at larger $r$. This behavior can likely be attributed to the singular behavior of $f^{(2)}$ near the threshold $r=1$.} Notably, the deviation reaches $8\%$ at $r = 0.85$, $17\%$ at $r=0.9$, and exceeds $40\%$ at $r = 0.95$. This is in stark contrast to the one-loop counterpart, where the difference between the asymptotic expression and the exact result is below $1\%$ even at $r=0.99$.  

%-----------------------------
%-----------------------------

\begin{figure}[!h]
\hspace{0cm}\includegraphics[width=0.45\textwidth]{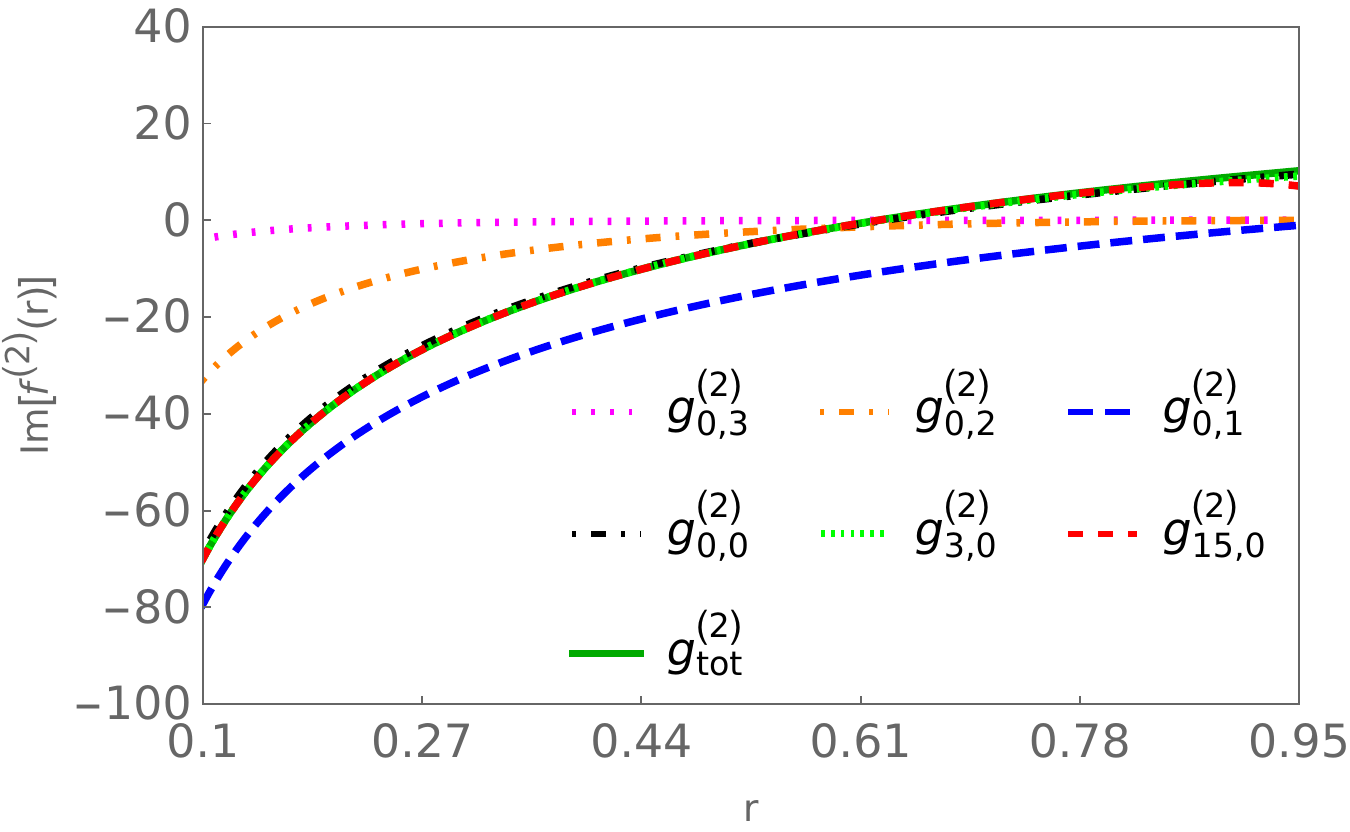}
\hspace{0cm}\includegraphics[width=0.45\textwidth]{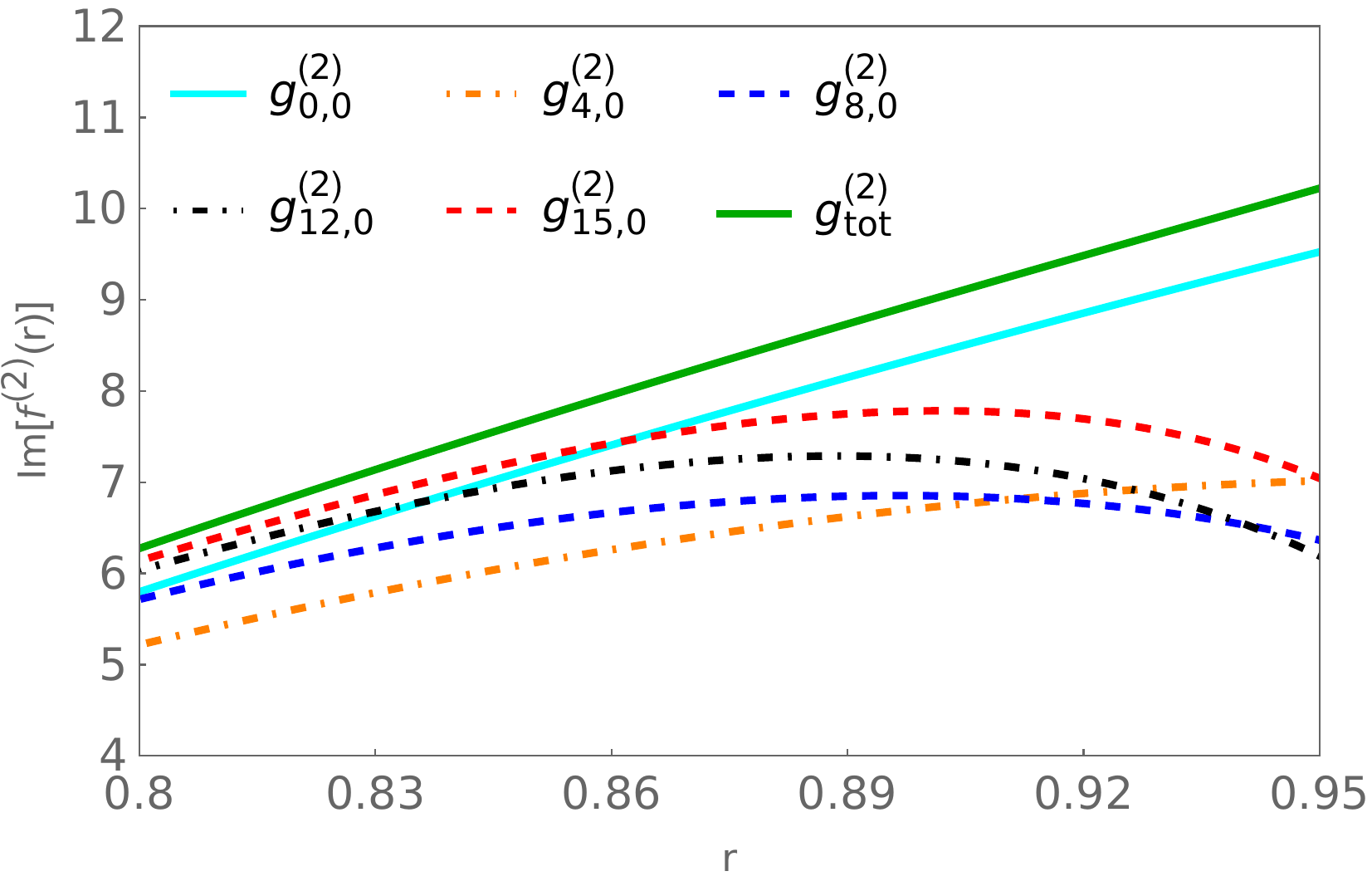}
\caption{\label{fig:loop2:im}
Comparison of the asymptotic expressions with the exact results for the imaginary part $f^{(2)}(r)$.}
\end{figure}

Similarly,  Fig.~\ref{fig:loop2:im} shows that the asymptotic expansion reproduces the imaginary part of $f^{(2)}$ accurately for $r < 0.8$, while the approximation visibly deteriorates when $r>0.8$.
The deviation reaches $6\%$ at $r = 0.85$, $13\%$ at $r = 0.9$, and exceeds $31\%$ at $r = 0.95$. For comparison, the deviation between the asymptotic expression and the exact results for the imaginary part of $f^{(1)}$ is $3\%$ at $r = 0.85$, $8\%$ at $r = 0.9$, and exceeds $20\%$ at $r = 0.95$.

%-----------------------------
%-----------------------------

Despite the poor convergence of the asymptotic expressions for $f^{(2)}$  and the imaginary part of $f^{(1)}$ near the threshold $s=16 m_c^2$ (i.e., for $r>0.8$), these expressions are still sufficient for making phenomenological predictions for the process $e^+e^-\to J/\psi+\eta_c$. In this context, $r$ is expected to be much less than $1$ to validate the applicability of the NRQCD factorization.

\label{sec-phe}
\section{Phenomenology}
Before the phenomenological discussion, we summarize the input parameters used in our analysis. The masses adopted are $M_{J/\psi} = 3.096916~\mathrm{GeV} $, $M_{\eta_c} = 2.9798~\mathrm{GeV} $~\cite{ParticleDataGroup:2024cfk}, and $m_c = 1.5 \pm 0.2~\mathrm{GeV} $. The NRQCD matrix elements are $\langle\mathcal{O}\rangle_{J/\psi} = 0.440~\mathrm{GeV}^3\ $and $\langle\mathcal{O}\rangle_{\eta_c} = 0.437~\mathrm{GeV}^3 $~\cite{Bodwin:2007fz}. The QED coupling constant is fixed at $\alpha(\sqrt{s} = 10.6~\mathrm{GeV}) = 1/130.9 $. 

We also use the following input values from~\cite{ParticleDataGroup:2024cfk}: $\alpha_s^{(n_f=5)}(M_Z) = 0.118 $ and $\overline{m}_c^{(n_f=3)}(\overline{m}_c) = 1.273~\mathrm{GeV} $. These values are converted to those with an active quark number $n_f=4$ using decoupling at flavor thresholds. The values of $\alpha_s $ and $\overline{m}_c$ at other energy scales are derived by solving the renormalization group equations with $n_f=4$ using the RunDec package~\cite{Chetyrkin:2000yt}. Additionally, the factorization scale is fixed at $\mu_\Lambda = 1~\mathrm{GeV}$.

%-----------------------------
\begin{figure}
\begin{center}
\hspace{0cm}\includegraphics[width=0.45\textwidth]{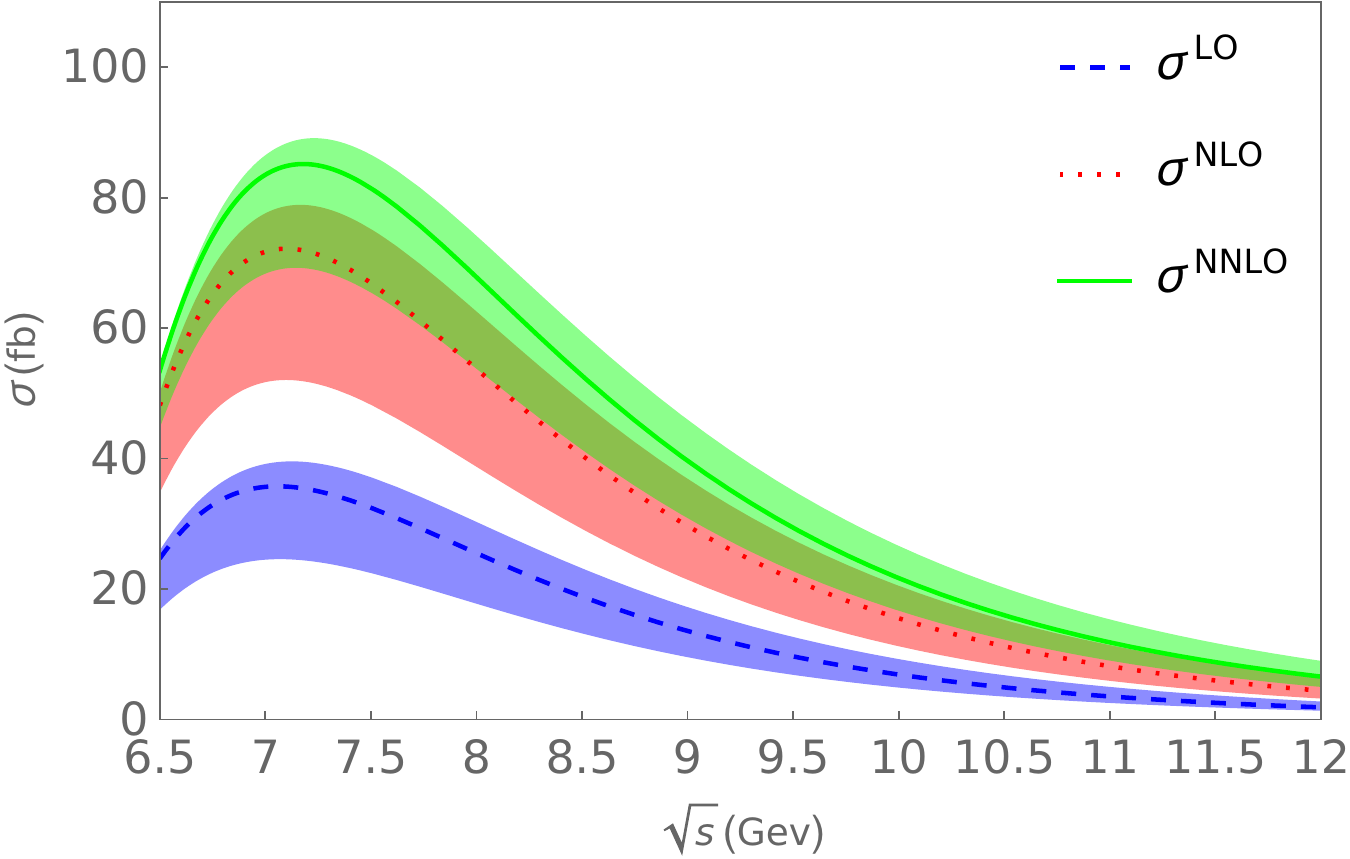}
\hspace{0cm}\includegraphics[width=0.45\textwidth]{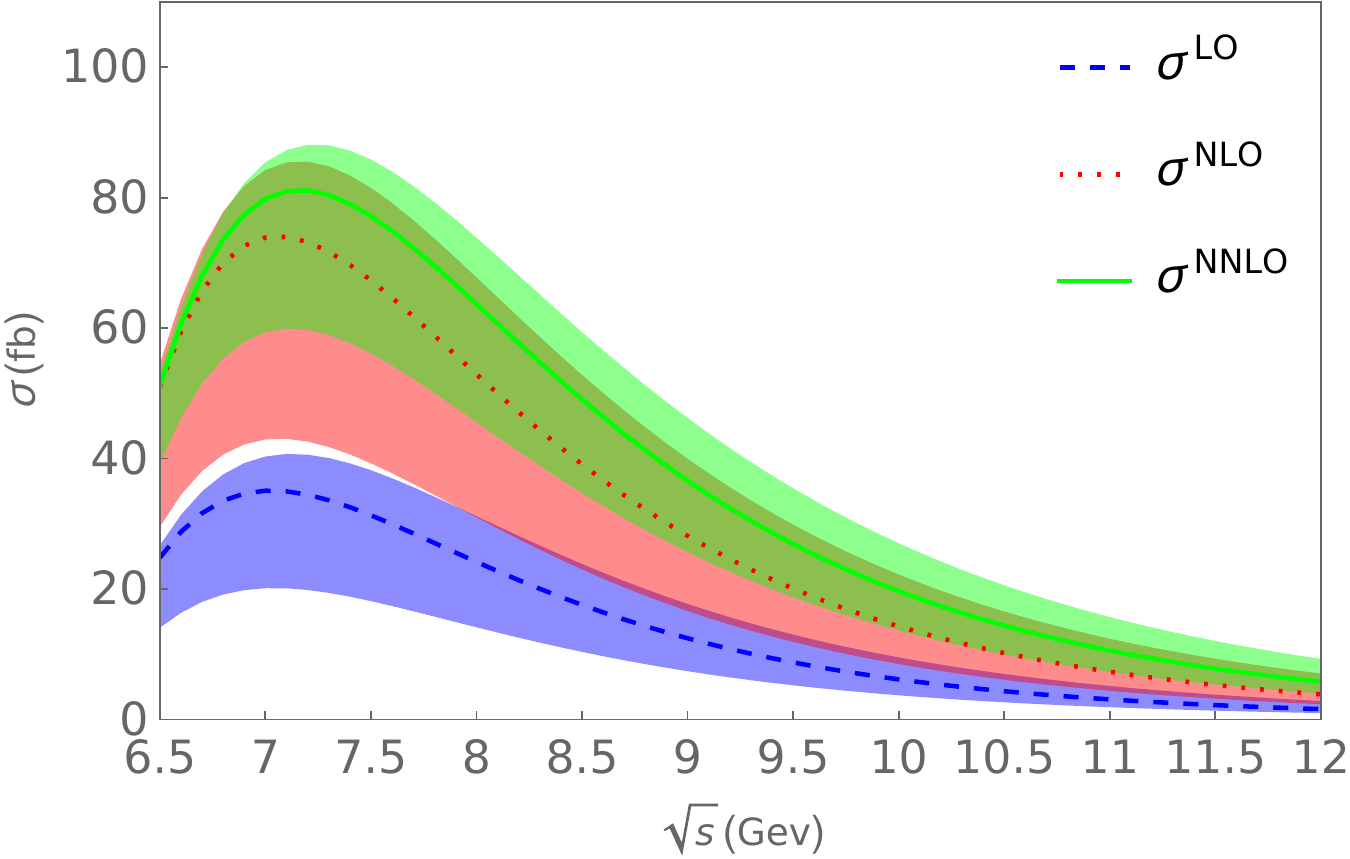}
\caption{\label{fig:cs}
The cross section as a function of $\sqrt{s}$ at various levels of accuracy in $\alpha_s$. The left panel  corresponds to predictions from the OS mass scheme, while right panel corresponds to predictions from the $\overline{\rm MS}$ mass scheme. For the OS mass scheme prediction, We take $m_c=1.5$ GeV. The shaded bands represent the uncertainty arising from varying $\mu_R$ from $3$ GeV to $\sqrt{s}$.}
\end{center}
\end{figure}

%-----------------------------
%For both mass schemes, the uncertainty in $\mu_R$ ranges from $2m_c = 3.$ GeV to $\sqrt{s}$, with the central value set at $\mu_R = \sqrt{s}/2$. In the on-shell mass scheme, the uncertainty due to the charm quark mass is considered by varying $m_c$ from 1.3 GeV to 1.7 GeV. In Fig.~\ref{fig:cs}, the shaded bands denote the uncertainty from $\mu_R$, while in Table~\ref{tab: 1}, the first column of data corresponds to the $\mu_R$ uncertainty and the second column represents the uncertainty from the charm quark mass.
%-----------------------------
%-------------------------
\begin{table*}[t]
\footnotesize
\setlength{\tabcolsep}{6pt}
\renewcommand{\arraystretch}{1.6}
\begin{center}
\caption{Theoretical predictions for the cross section (in fb). For the OS mass scheme, the uncertainties arise from two sources: varying $\mu_R$ from $3$ GeV to $\sqrt{s}$, and varying the charm quark pole mass from $1.3$ GeV to $1.7$ GeV. For the $\overline{\rm{MS}}$ mass scheme, the uncertainty is associated with varying $\mu_R$ from $3$ GeV to $\sqrt{s}$.}
\label{tab: 1}
\begin{tabular}{ccccccccc}
\hline
\multirow{2}{*}{$\sqrt{s}(\mathrm{GeV})$} & \multicolumn{3}{c}{OS scheme} & \multicolumn{3}{c}{$\overline{\rm MS}$ scheme} \\ \cline{2-7}
 & LO & NLO & NNLO & LO & NLO & NNLO  \\ \hline \hline
10.58 & $4.7^{+1.8+1.5}_{-1.3-1.0}$ & $10.7^{+3.9+4.6}_{-2.9-2.9}$ & $15.2^{+4.1+7.6}_{-3.6-4.6}$&$4.1^{+2.5}_{-1.6} $&$9.7^{+6.1}_{-3.9} $&  $13.7^{+5.7}_{-4.7}$ \\  \hline
\end{tabular}
\end{center}
\end{table*}
%-------------------------

In Fig.~\ref{fig:cs}, we present the cross section as a function of $\sqrt{s}$ and make several key observations. First, the cross section calculated using the OS mass scheme is slightly larger than that obtained from the $\overline{\rm{MS}}$ mass scheme. The shapes of the curves from both mass schemes are quite similar. Specifically, the cross section initially increases and then decreases with the increase of $\sqrt{s}$, reaching its maximum value at $\sqrt{s}$ close to $7$ GeV. It is worth noting that while the NRQCD factorization is not strictly valid at the lower end of the $\sqrt{s}$ range and therefore fail to provide fully reliable predictions, it still offers a reasonable theoretical estimate in this region.

Second, we observe that the perturbative corrections are substantial. The $\mathcal{O}(\alpha_s^2)$ corrections are slightly smaller than the $\mathcal{O}(\alpha_s)$  corrections, particularly at lower values of $\sqrt{s}$.

Third, we find that the $\mu_R$ dependence of the predictions from the $\overline{\rm{MS}}$ mass scheme is slightly larger than that from the OS mass scheme.~\footnote{Recall that, as explained in Sec.~\ref{sec-mass-scheme}, we have fixed the scale $\mu_m=\overline{m}_c(\overline{m}_c)$ when converting the pole mass—which arises from perturbative matching in the charmonium rest frame and appears explicitly in Eq.~\eqref{cross:section}—into the $\overline{\rm{MS}}$ mass.} This is somewhat counterintuitive and contrary to our initial expectations. In fact, a substantial portion of the uncertainty in the predictions from the $\overline{\rm{MS}}$ mass scheme stems from the conversion of the pole mass $m_c$ in \eqref{eq:LO}, which arises due to the helicity flip. In contrast, for helicity-conserved processes, the predictions from the $\overline{\rm{MS}}$ mass scheme are expected smaller than those from the OS scheme.

In Table~\ref{tab: 1}, we present the cross section for the process $e^+e^-\to J/\psi+\eta_c$ at a specific value of $\sqrt{s}=10.58$ GeV, calculated using both the OS mass scheme and the $\overline{\rm MS}$ mass scheme. The table includes uncertainties arising from the renormalization scale. Additionally, we account for the uncertainty in the charm quark mass within the OS mass scheme prediction by varying $m_c$ from $1.3$ to $1.7$ GeV. 

The perturbative corrections are notably large. In the OS scheme the $\mathcal{O}(\alpha_s)$ correction amounts to about $6$ fb, which exceeds the LO contribution, while the $\mathcal{O}(\alpha_s^2)$ term is approximately $4.5$ fb—smaller than the $\mathcal{O}(\alpha_s)$ term but still substantial. This pattern suggests a slow perturbative convergence. The corresponding corrections in the $\overline{\text{MS}}$ scheme are somewhat smaller.

It is found that the uncertainty stemming from the charm quark mass is comparable to, or even larger than, that from the renormalization scale, thus constituting one of the main sources of uncertainty for the OS mass scheme prediction. The cross sections obtained from the OS mass scheme are consistent with those from the $\overline{\rm MS}$ scheme within the uncertainties.

At $\sqrt{s} = 10.58$ GeV, our predictions agree with the BABAR and Belle measurements within errors. 
It is worth noting that, with the formulas and expressions provided in the attached file, one can easily calculate the cross sections at other CM energies.

\label{Summary}
\section{Summary} 
We present NNLO QCD corrections to the SDC for $e^+e^-\to J/\psi+\eta_c$ in asymptotic expansions up to $r^{15}$. These expressions significantly facilitate both theoretical and phenomenological studies.
We provide phenomenological predictions for the cross sections in both the OS mass scheme and the $\overline{\rm MS}$ mass scheme up to NNLO.  
We estimate theoretical uncertainties by varying $\mu_R$ and charm quark pole mass in the OS scheme, while varying only $\mu_R$ in the $\overline{\rm MS}$ scheme.
The $\mu_R$ uncertainty in the $\overline{\rm MS}$ mass scheme is slightly larger than that from the OS scheme, primarily due to the helicity flip. 
%However, when accounting for the uncertainty from the charm quark pole mass, the total uncertainty in the OS mass scheme becomes larger than that in the $\overline{\rm MS}$ mass scheme.  
Finally our theoretical results are compatible with the available experimental data.
%------------------------------

\section{Acknowledgments} 
The work of C. Li. and W.-L. S. is supported by the
National Natural Science Foundation of China under Grants
No. 12375079, and the Natural Science
Foundation of ChongQing under Grant No. CSTB2023
NSCQ-MSX0132. 
%---------------------
The work of X.-D. H. is supported by the
National Natural Science Foundation of China under Grants
No. 12505097, and the Foundation of Chongqing Normal University under Grant No. 24XLB015.

%----------------------------------------

\end{document}